\documentclass[aps,pra,twocolumn,superscriptaddress]{revtex4}

\usepackage{epsfig}
\usepackage{graphicx}
\usepackage{enumitem, hyperref}
\usepackage{color}

\begin{document}

\title{Phase Control of Squeezed Vacuum States of Light in Gravitational Wave Detectors}

\author{K.~L. Dooley}
\email[Please address correspondence to: ]{kate.dooley@aei.mpg.de}
\author{E. Schreiber}
\author{H. Vahlbruch}
\author{C. Affeldt}
\author{J.~R. Leong}
\author{H. Wittel}
\author{H. Grote}
\affiliation{Max-Planck-Institut f\"ur Gravitationsphysik (Albert Einstein Institut) and
Leibniz Universit\"at Hannover, Callinstr.\,38, 30167 Hannover, Germany}


\date{\today}

\begin{abstract}
Quantum noise will be the dominant noise source for the advanced laser
interferometric gravitational wave detectors currently under
construction. Squeezing-enhanced laser interferometers have been
recently demonstrated as a viable technique to reduce quantum
noise. We propose two new methods of generating an error signal for
matching the longitudinal phase of squeezed vacuum states of light to
the phase of the laser interferometer output field. Both provide a
superior signal to the one used in previous demonstrations of
squeezing applied to a gravitational-wave detector. We demonstrate
that the new signals are less sensitive to misalignments and higher
order modes, and result in an improved stability of the squeezing
level. The new signals also offer the potential of reducing the
overall rms phase noise and optical losses, each of which would
contribute to achieving a higher level of squeezing. The new error
signals are a pivotal development towards realizing the goal of 6\,dB
and more of squeezing in advanced detectors and beyond.
\end{abstract}


\maketitle

\section{Introduction}
The dominant broadband noise source for the advanced laser
interferometric gravitational wave (GW) detectors will be quantum noise
\cite{Harry2010Advanced, Somiya2012Detector,
  TheVirgoCollaboration2009Advanced}. The classical method for
reducing quantum noise at shot-noise-limited frequencies is to
increase the laser power. Higher laser power, however, introduces many
technical challenges arising from laser light absorption and
subsequent heating of the optics. Another approach to reduce quantum
noise is to inject squeezed states of light into the interferometer's
anti-symmetric port, a technique which reduces the measurement
uncertainty in the readout signal
\cite{Caves1981QuantumMechanical}. Rapid advances in both squeezing
technology and laser interferometer development in the last decade
resulted in the first demonstrations of this quantum noise reduction
technique on current interferometric GW detectors in 2010 at GEO\,600
\cite{2011Gravitational} and in 2011 at LIGO Hanford
\cite{Collaboration2013Enhancing}.

GEO\,600 is carrying out the first long-term study of incorporating
squeezed states of light in a GW detector. Results include
demonstration of a squeezing duty cycle of 90\% with mean detected
squeezing of 2.0\,dB during an 11~month data collection period in 2012
\cite{Grote2013First}. Continued work since then has resulted in an
increase of the observed squeezing level up to a maximum of 3.7\,dB to
date and a continued high duty cycle of 85\%
\cite{Affeldt2014Advanced}. This study has demonstrated the readiness
of squeezed states of light as a permanent application for increasing
the astrophysical reach of GW detectors. Projects such as Advanced
LIGO and Advanced Virgo are now making plans to incorporate squeezing
as an early instrumental upgrade.

The limits to the level of non-classical noise reduction that can be
achieved depend on the following four variables: degree of generated
squeezing, optical losses (including beam alignment and
mode-matching), phase noise, and noises in the squeezing frequency
band other than shot noise. This paper focuses on new techniques
developed, implemented, and analyzed at GEO\,600 which serve to reduce
phase noise.

\begin{figure}
\begin{centering}
\includegraphics[width=0.6\columnwidth]{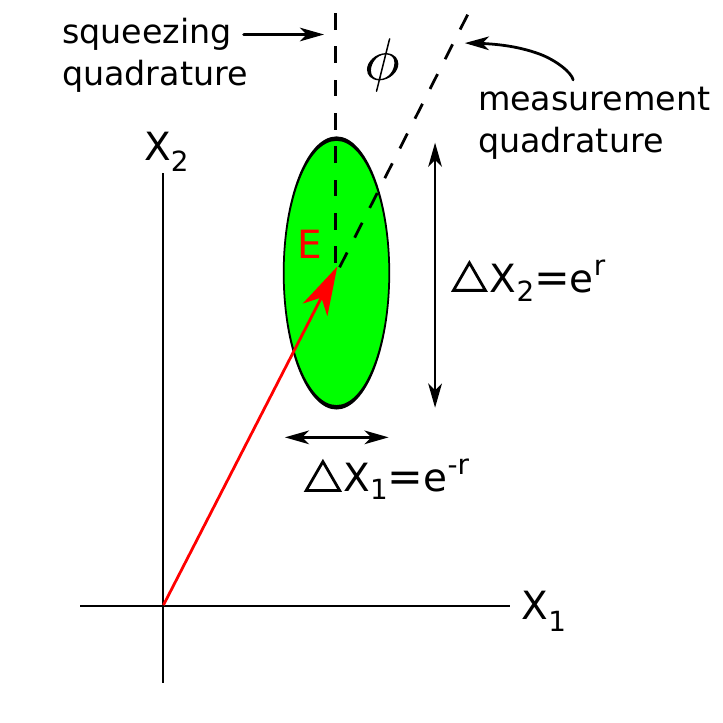} 
\caption{Phasor diagram of a squeezed state
  ($\Delta X_1 \Delta X_2 \geq~1$). The factor $r$ describes the
  degree of squeezing and anti-squeezing for a pure state and $\phi$
  describes the mismatch in angle between the squeezing and
  measurement quadratures. For application in a GW detector, phase
  squeezing is injected to the anti-symmetric port.}
\label{fig:cartoon}
\end{centering}
\end{figure}

Phase noise refers to any root-mean-square (rms) difference between
the angle of the squeezing ellipse and the angle of the measurement
quadrature of the interferometer as depicted in
Fig.~\ref{fig:cartoon}. The degree of measurable squeezing and
anti-squeezing is reduced for an rms phase noise of $\phi\ll\pi$ as
follows:
\begin{eqnarray}
V_s^{\prime} =& V_s \cos^2{\phi} + V_a \sin^2{\phi} \\
V_a^{\prime} =& V_a \cos^2{\phi} + V_s \sin^2{\phi} 
\end{eqnarray}
where $V_s$ and $V_a$ are the variances of the squeezed and
anti-squeezed states, respectively, before including the effect of
phase noise. 

Phase noise, also called `quadrature fluctuations' or `squeezing angle
jitter', is one of the limits to quantum noise reduction that already
affects today's squeezing enhanced interferometers. During regular
squeezing operation at GEO\,600, there are approximately 35\,mrad rms
phase noise. With optical losses of about 40\%, this phase noise
reduces the observed squeezing level by a few tenths of a dB compared
to no phase noise at all. In the extreme case, too much phase noise
can even result in anti-squeezing, as was observed during the
squeezing experiment at LIGO Hanford when a high non-linear gain was
intentionally used \cite{Dwyer2013Squeezed}.

As optical losses are lowered, phase noise becomes more
critical. Anti-squeezing grows larger and its projection onto the
squeezed state thus also grows larger for a given angle. To achieve
6\,dB of squeezing as is intended for advanced detectors, phase noise
must be no more than 10\,mrad rms if optical losses are 25\%. However,
with a push to only slightly lower optical losses, such as 20\%, as
much as 30\,mrad phase noise can be tolerated. Third generation
detectors, which have goals of 10\,dB of squeezing
\cite{Hild2011Sensitivity}, will require that phase noise be at most
only a couple mrad rms. Here, a critical boundary is that losses of
10\% would already require there be no phase noise at all.

Both static mismatch and relative motion at all frequencies between
the squeezing and measurement quadratures contribute to the rms phase
noise. Temperature-induced path length fluctuations, swinging
suspended optics, and phase modulation from radio frequency (RF)
sidebands account for some of the sources of phase noise. Calculations
of these and other effects are presented in
Ref.~\cite{Dwyer2013Squeezed}. Described in the context of a phase
noise sensing and control system, the total phase noise can be grouped
into contributions from four frequency bands:
\begin{itemize}
\item DC: lock point errors
\item in-loop frequencies: integrated rms within the control band
\item audio frequencies: phase noise outside of the control band 
\item radio frequencies: RF sidebands used for interferometer control
  create phase noise on the GW carrier
\end{itemize}

Efforts to minimize phase noise at GEO\,600 are three-fold. First,
steps are taken to build an intrinsically quiet squeezing source to
limit the amount of fluctuations of the squeezing ellipse at its
generation. This includes considerations in the mechanical design of
the optical parametric amplifier (OPA) as well as the implementation
of a pump phase control loop for stabilizing the squeezing angle when
it is created. Overall, the GEO\,600 OPA produces a squeezing ellipse
with 9\,mrad rms phase noise \cite{Khalaidovski2011Beyond}. Second,
this stable squeezed field is in turn stabilized with respect to the
GW carrier at the interferometer output port using coherent control
sidebands (CCSBs) on the squeezed
field~\cite{Vahlbruch2006Coherent}. Third, drift of the squeezing
angle which cannot be sensed properly by the coherent control loop is
counteracted at frequencies $<~10$~mHz through a noise locking
technique \cite{McKenzie2005Quantum} to maximise the strain
sensitivity. The combination of the noise lock with coherent control
is new, and was pivotal for long-term squeezing \cite{Grote2013First}.

Historically, both at GEO\,600 and at LIGO Hanford, the phase error
signal was derived from the beat between the squeezer CCSBs and the
interferometer carrier light at a 1\,\% pick-off mirror before the
output mode cleaning cavity (OMC) \cite{2011Gravitational,
  Collaboration2013Enhancing}. This signal has both susceptibility to
lock point errors due to higher order modes (HOMs) and has a limited
signal to noise ratio (SNR). We present our study and experimental
demonstration at GEO\,600 of the advantages of two alternative
techniques of generating coherent control phase error signals.

This paper is organized as follows: Section~\ref{sec:experiment}
provides an overview of the experimental setup;
Section~\ref{sec:signals} introduces the three phase error signals and
discusses their respective merits and drawbacks;
Section~\ref{sec:results} presents our experimental results; and
Section~\ref{sec:discussion} discusses implications for the design of
future squeezing-enhanced GW detectors. The paper finishes with a
summary in Section~\ref{sec:conclusion}.

\section{Experimental setup}
\label{sec:experiment}

\begin{figure*}
\begin{centering}
\includegraphics[width=2\columnwidth]{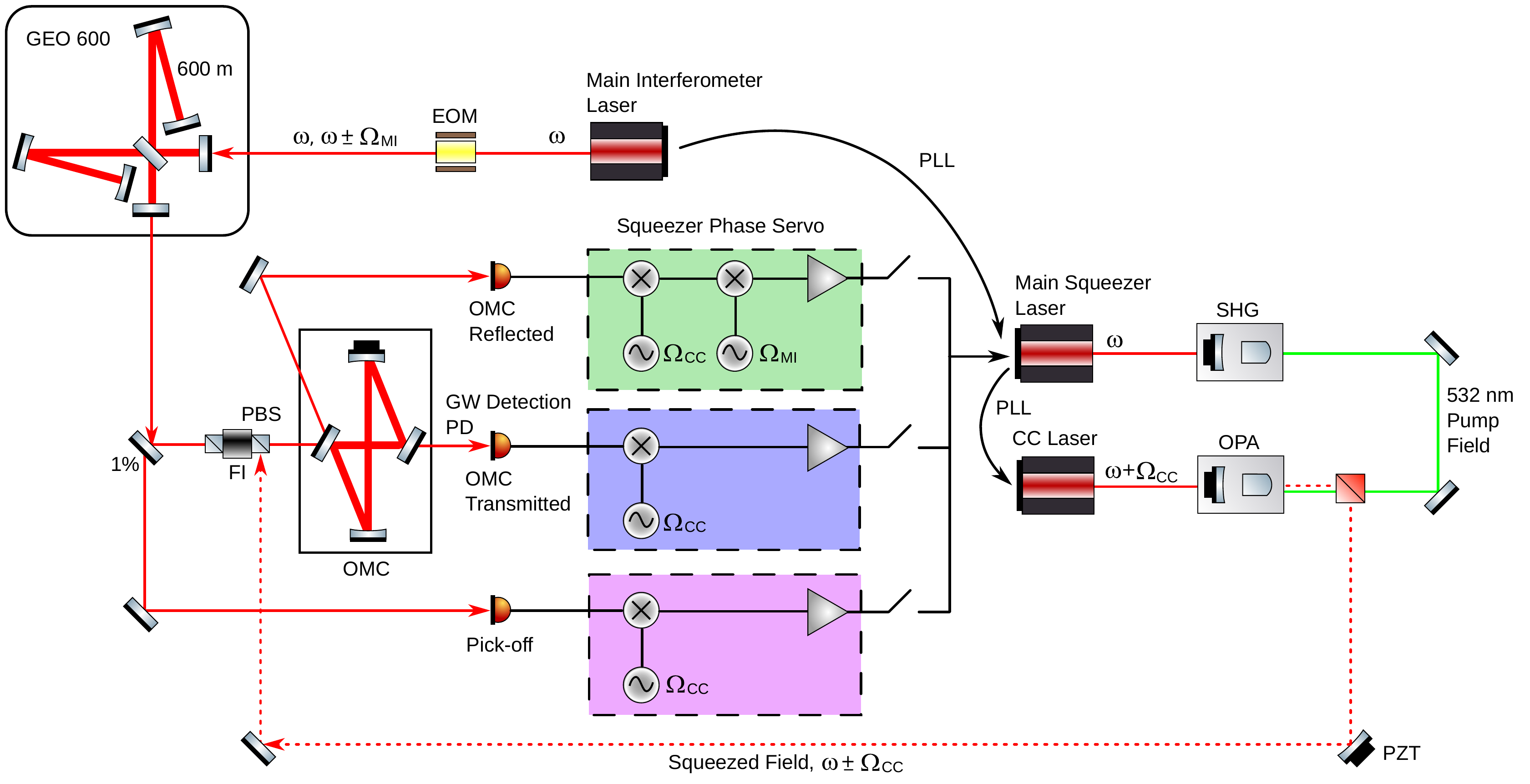} 
\caption{Schematic of the GEO\,600 interferometer together with the
  squeezing source. The three methods presented in this article for
  generating a squeezer phase error signal are highlighted: OMC
  reflected, OMC transmitted, and pick-off. Only one path is used at a
  time and the control signal is fed back to the error point of the
  PLL which locks the main squeezer laser to the interferometer main
  laser. Complete phase control is accomplished through the following
  two additional stages: locking of the green pump beam phase to the
  OPA and a noiselock loop.}
\label{fig:overview}
\end{centering}
\end{figure*}

Figure~\ref{fig:overview} shows a schematic of the GEO\,600
interferometer together with the squeezing source and highlights how
the three different phase error signals are generated. The optical
layout of GEO\,600 is depicted in the upper left corner. GEO\,600 is a
power- and signal-recycled Michelson interferometer with folded arms
within 600\,m long beam tubes \cite{Grote2010GEO}. Gravitational waves
phase modulate the carrier light in the Michelson arms and the
resulting audio frequency sidebands are coupled out to the
anti-symmetric port of the interferometer. The Michelson is operated
with a small dark fringe offset so that some carrier light leaks to
the anti-symmetric port and serves as a local oscillator for the GW
sidebands \cite{Hild2009DCreadout}. Beam directing and mode-matching
optics send this light to the OMC to filter out HOMs and RF control
sidebands. The GW signal is encoded as power variations of the light
transmitted through the OMC and is detected by an in-vacuum
photodetector (PD).

The GEO\,600 squeezed light source is installed on an in-air table
next to the vacuum tank containing the OMC and readout PD. A series of
steering mirrors directs the squeezed field to the open port of the
polarizing beam splitter (PBS) of the in-vacuum Faraday isolator (FI)
in the interferometer output chain. Located directly in front of the
OMC, the PBS directs the squeezed field backwards through the FI,
rotating its polarization to match that of the interferometer
carrier. The squeezed vacuum is then reflected off of the over-coupled
signal recycling cavity (SRC) and joins the GW local oscillator field
on the detection PD.

For the generation and control of squeezed vacuum states, three
phase-locked lasers are used (only two of which are indicated in
Fig.~\ref{fig:overview}). A fraction of the main squeezer laser at
1064\,nm is frequency doubled in a second-harmonic generator (SHG)
which provides the required pump field at 532\,nm for the non-linear
squeezing resonator. One of the control lasers (not shown) locks the
OPA length. The other control laser (labeled `CC') is injected into
the locked OPA to stabilize the angle of the squeezing ellipse with a
bandwidth of 7~kHz. This process generates the CCSBs which have a
stable phase with respect to the correlated audio sidebands
\cite{Vahlbruch2006Coherent}. The CCSBs thus serve as a marker of the
squeezing angle and co-propagate with the squeezed field. The CCSB
frequency, $f_{\mbox{CC}}=15.2$~MHz, is chosen so as to be
anti-resonant in the GEO SRC. A detailed description of the GEO\,600
squeezer is found in Ref.~\cite{Vahlbruch2010GEO}.

The three methods for generating a squeezer phase error signal are
featured in the center of Fig.~\ref{fig:overview} and will be
discussed in detail in the next sections. The error signal is fed back
to change the frequency of the squeezer main laser, which acts as a
phase actuator with a $1/f$ response. A gain is selected to give a
bandwidth of about 2~kHz and some filters to provide additional
suppression below 30~Hz are also included. In addition, a calibration
line is injected at 6500~Hz by actuating longitudinally via a PZT
mounted on a steering mirror in the squeezer path.

\section{Phase error signal}
\label{sec:signals}
The squeezer phase error signal is contained in the beat between the
TEM$_{00}$ modes of fields which carry phase information about the GW
carrier and the squeezed vacuum, respectively. Potential signals are
derived not only from various choices of reference fields, but also
from a selection of ports where these fields are available. We
evaluate the following three methods of generating a squeezer phase
error signal with respect to the GW measurement quadrature:
\begin{itemize}
\item CCSBs vs. carrier at pick-off port
\item CCSBs vs. Michelson sidebands in OMC reflected
\item CCSBs vs. carrier in OMC transmitted
\end{itemize}
We refer to the signals as `pick-off', `OMC reflected', and `OMC
transmitted', respectively, where the latter two are the new
alternative techniques.

The squeezer CCSBs are used as a reference of the squeezer phase in
all scenarios because the squeezed field itself cannot be used to
generate an error signal. It contains only several photons per second
\cite{Mehmet2010Observation} and such a low amplitude field cannot be
measured directly on a photodetector. In addition, the interferometer
output field contains not only the local oscillator for the GW signal,
but also the interferometer RF control sidebands which serve as a
phase reference of the carrier light. Both the carrier and the
Michelson (MI) sidebands are thus candidates for representing the
squeezing quadrature to generate a phase signal. One feature of
GEO\,600's MI sidebands ($f_{\mbox{MI}}=14.9$~MHz) is that they are
spatially cleaner at the output port than the carrier \footnote{The
  Schnupp asymmetry means the MI sidebands don't see the imperfect
  mirrors as many times as the recycled carrier field.}.

If present, higher order modes play a central role in the quality of
the error signals both through increasing the shot noise but not the
signal and through creating an offset to the lock point of the
loop. We define \emph{lock point errors} as non-intentional
contamination of the proper phase signal with false information which
pushes the system away from the nominal operating point. Because these
offsets originate on the sensor and are thus in-loop, they cannot be
suppressed by the loop. Intrinsic HOM content of either the local
oscillator or reference fields, mode-mismatch of the fields, and beam
misalignment are all relevant factors for creating lock point
errors. Detailed calculations and additional discussion are presented
in Ref.~\cite{Oelker2014Squeezed}.


The SNR for a given phase error signal is defined as:
\begin{equation}
\mbox{SNR} \propto \frac{E_1E_2}{\sqrt{P}}
\end{equation}
where $E_1$ and $E_2$ are the amplitudes of the signal fields and $P$
is the total power on the PD, which is valid as long as the sensor is
shot-noise-limited. A high SNR allows in-band phase noise to be
reduced: for a given bandwidth of a sensor-noise-limited servo, there
is a linear relationship between noise floor reduction and in-band rms
noise reduction. The most pertinent factors to consider for achieving
high SNR are the existence of HOMs and port selection. HOMs contribute
to the total power (i.e. noise) but not to the signal and can be
reduced through the use of mode cleaning cavities. Additionally,
improvements in SNR can come from selecting ports that have a
favorable transmission of the signal fields.


\begin{table}
\centering
\caption{Schemes studied for generating a squeezer phase error
  signal. Signal-to-noise ratios are compared as well as the
  likelihood of lock point errors as determined by the quantity of
  HOMs in the signal fields. Field amplitudes and the SNR are
  normalized to 1.}
\begin{tabular}{l l l l}
\hline
         & Pick-off (1\%) & OMC refl. & OMC trans. \\
\hline
fields & [CCSB, carrier] & [CCSB, MISB] & [CCSB, carrier] \\
amplitudes & [0.1, 0.1] & [1.0, 0.3] & [0.1, 1.0] \\
frequency & 15.2\,MHz & 300\,kHz & 15.2\,MHz \\
total power & 0.3\,mW & 24\,mW & 6\,mW \\
SNR & 1/3 & 1  & 2/3 \\
HOMs & $80\%$ & $<1\%$ & $\ll1\%$ \\
\hline
\end{tabular}
\label{tab:signals}
\end{table}

The pick-off before the OMC is a 1\% power transmissive mirror which
is nominally in place to extract alignment signals for the MI
interferometer. All of the light experiences this 1\%
transmission. The OMC rejects a significant fraction of the CCSBs and
MI sidebands in addition to HOMs. Thus, all fields except the carrier
TEM$_{00}$ are in the OMC reflected port. The carrier is entirely
transmitted. The OMC finesse of 150 is sufficiently low, however, such
that the CCSBs and MI sidebands do have a power transmission of
1\%. The CCSBs are thus available at all three ports, although to
varying extents, which plays a role in the available SNR. Furthermore,
although the MI sidebands have intrinsically less HOM content than the
carrier, the carrier is stripped of its HOMs by the OMC, making it a
promising signal at this transmission port.

Based on the pick-off fractions at each port and measurements of the
power in the GEO\,600 output beam, the SNRs of the three signals are
computed and displayed in Table~\ref{tab:signals}. Of the
approximately 30~mW in the GEO\,600 output port beam, about 6~mW are
carrier light in the TEM$_{00}$ mode and approximately 0.6~mW are
Michelson sidebands. The remainder are HOMs, predominantly at the
carrier frequency. The highest of the field amplitudes and SNRs are
normalized to 1 to allow easier comparison of signals. The fraction of
the total power in each reference field that is made of HOMs is also
presented. For the pick-off signal, this is based on power
measurements. For the OMC reflected signal, where HOMs of only the
Michelson sideband light are of interest, OMC mode scans using two
different sideband powers provides the upper limit of 1\%. The
percentage of carrier HOMs in the OMC transmitted light is computed
based on the OMC g-factor and finesse and the mode content of the
output port beam.

In terms of both SNR and susceptibility to lock point errors, we find
that the pick-off signal is the worst option. For GEO\,600 the signal
in OMC reflected has a 3-fold higher SNR and less HOM content. The
signal in OMC transmission has a 2-fold higher SNR and almost no HOM
content. The reduction of HOMs in the signal is a very important
feature in that it holds the promise of eliminating lock point errors,
a critical problem that was encountered in the LIGO squeezing
demonstration \cite{Dwyer2013Squeezed} and to a lesser extent at
GEO\,600 due to the addition of a squeezer alignment system
\cite{SchreiberAlignment}.

\section{Experimental results}
\label{sec:results}

We experimentally demonstrate the reduced lock point errors of the two
new phase error signals using the squeezing-enhanced GEO\,600
detector. We built and installed the appropriate electronics and
photodiodes as depicted in Figure~\ref{fig:overview} and commissioned
each of the control loops. As the signals cannot be used
simultaneously, results were obtained by switching consecutively from
one signal to the next on a seismically quiet evening during a single
lock stretch to ensure the fairest possible comparisons. Twenty-minute
long data stretches were acquired for each signal and the
reproducability of the results were verified by repeating the entire
experiment on several occasions. The auto-alignment system for the
squeezer was used and the squeezer tuned to achieve 3.2\,dB, the
highest squeezing level possible at the time.

\begin{figure}
\begin{centering}
\includegraphics[width=\columnwidth]{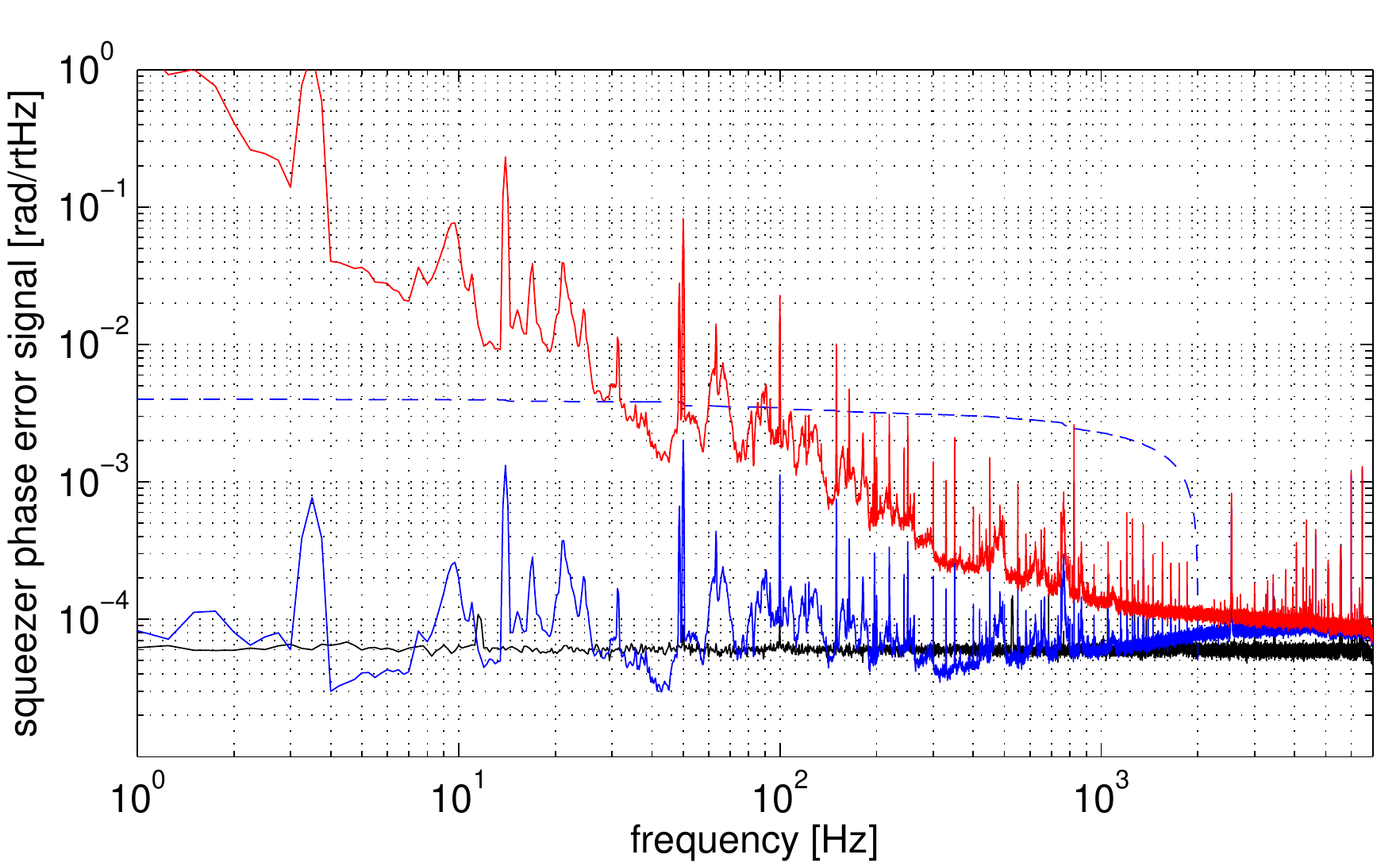}
\caption{Phase noise detected by the beat of the CCSBs and the carrier
  light in OMC transmission. The in-loop error signal (blue) is
  suppressed to or below sensor noise (black). The free-running phase
  noise (red) is computed from the in-loop spectrum and the known loop
  shape. The dashed line shows the integrated rms within the control
  band of 2\,kHz. The in-loop phase noise is 4\,mrad rms. Spectra are
  created from a digitally-acquired 240 sec long time series with
  0.25\,Hz binwidth and 120 averages. A calibration line is injected
  at 6500\,Hz.}
\label{fig:signals}
\end{centering}
\end{figure}

\begin{figure}
\begin{centering}
\includegraphics[width=\columnwidth]{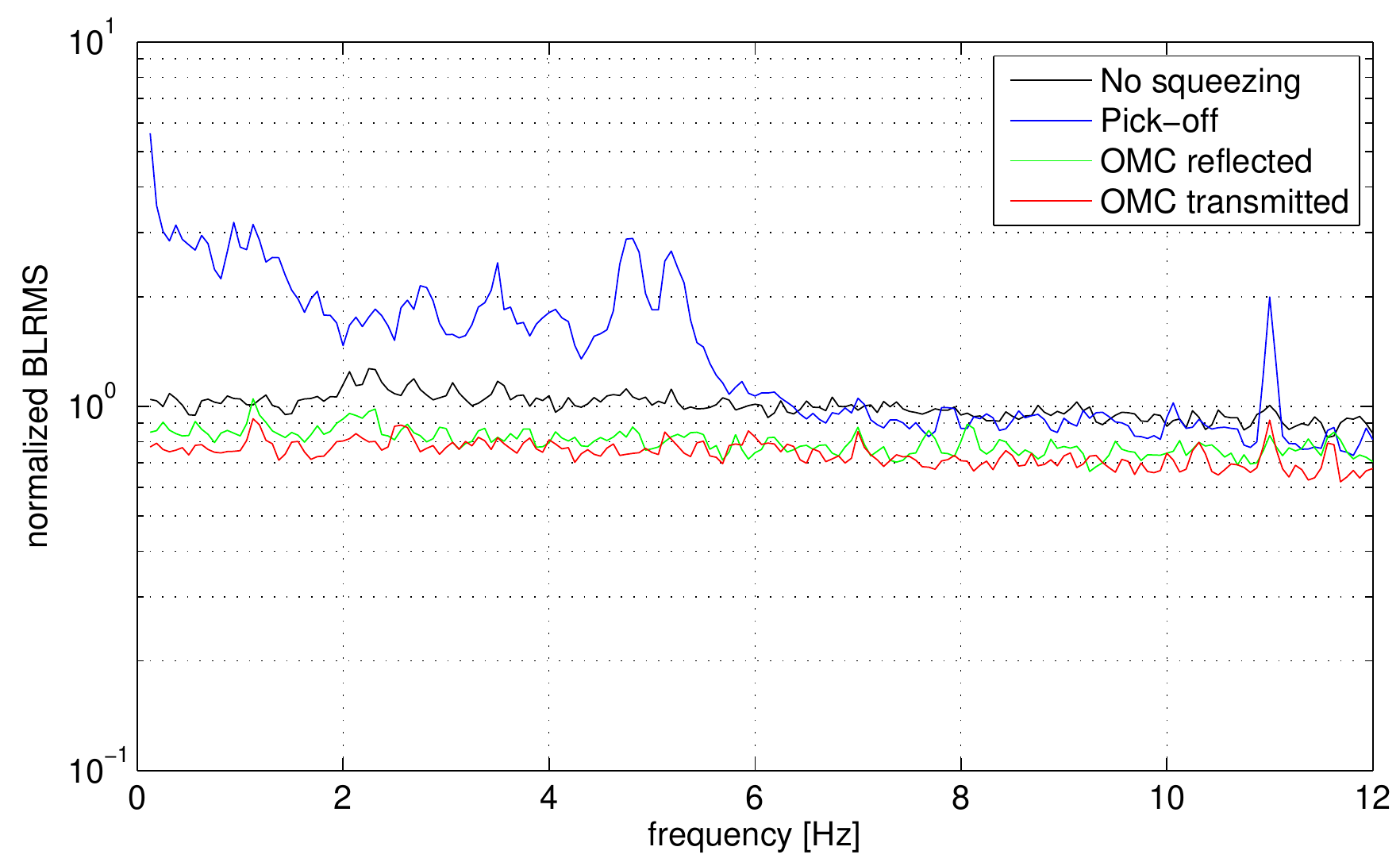}
\caption{Amplitude spectra of the band-limited-rms of strain
  sensitivity in a shot-noise-limited frequency band from 4\,kHz to
  5\,kHz. Stability of the squeezing level when using the OMC
  transmission and OMC reflection signals is demonstrated. The
  increase in fluctuations of the squeezing level when the standard
  pick-off signal is used is due to lock point errors from HOMs. All
  spectra are normalized to the mean of the non-squeezed spectra and
  the relative levels indicate the amount of squeezing present.}
\label{fig:blrms}
\end{centering}
\end{figure}

Figure~\ref{fig:signals} shows a representative example of the free
running, in-loop, and sensing noise spectra of the squeezer phase
error signal. The free-running spectrum was computed based on the
measured in-loop spectrum and a model of the open loop transfer
function. The sensing noise spectrum is taken when a shutter in the
optical path of the squeezed field is closed and thus no squeezing
applied. All spectra are calibrated to $\mbox{rad}/\sqrt{\mbox{Hz}}$
using the measured peak-to-peak amplitude $V_{pp}$ of the free-running
error signal. The calibration factor is $1/(V_{pp}/2)$.

This particular example is from the OMC transmission
signal. Equivalent spectra from the pick-off and OMC reflection
signals are nearly identical and feature only different sensing noise
floor levels. Unfortunately, the experimental setup does not fully
reflect the ideal scenario for which SNRs were calculated in
Table~\ref{tab:signals} such that a direct comparison of noise floor
levels to predicted SNR cannot be made. For instance, none of the
sensors are shot-noise-limited; electronics noise sits close below
sensor noise in all scenarios. In the case of OMC transmission, this
extra noise has been tracked to RF pick-up problems with the
photodiode setup. Furthermore, in generating the signal in OMC
reflection, only some fraction of the available light is used.

The magnitude of the effect of extra electronics noise on the total
in-loop phase noise is of interest. A measurement of dark noise
indicates that the shot-noise-limit is 25\% below sensor noise. The
in-loop phase noise contribution is computed from the quadrature sum
of the error signal and sensor noise starting at the loop's unity gain
frequency of 2\,kHz. It is 4\,mrad rms, but could be reduced to 3\,mrad
rms should the RF pick-up noise in this setup be reduced.

Figure~\ref{fig:blrms} highlights the main result of this paper, which
is that the OMC transmission and OMC reflection phase error signals
eliminate lock point errors and create a squeezing level which is
stationary in time. The band-limited rms (BLRMS) spectra for a
shot-noise-limited region between 4\,kHz and 5\,kHz of the
GEO\,600 strain sensitivity is plotted for a non-squeezed time as well
as squeezed times when each of the various phase error signals were
used. If the total rms phase noise changes in time, then there will be
fluctuations of the squeezing level and therefore of the
shot-noise-limited strain sensitivity. There are both linear and
quadratic couplings of the squeezing angle fluctuations to the
frequencies in this spectrum, depending on the frequency content and
amplitude of the fluctuations.

Notably, the spectrum of the shot noise BLRMS when the OMC
transmission and OMC reflection signals are used is no different than
that from the non-squeezed time. Only the DC level differs, which is
the desired effect of squeezing. Furthermore, this demonstrates that
the total rms phase noise from in-band and RF frequencies is very
stable on these time scales. The effect of lock point errors from the
pick-off signal is largely observed at frequencies below 6~Hz where
the suspended output optics are swinging and altering the spatial
distribution of HOMs on the sensor. These high levels of phase noise
are not so frequent, so the effect on the average squeezing level is
small. Nonetheless, stationarity of the data is of great importance
for gravitational wave searches.

\begin{table}
\centering
\caption{Phase noises from known sources when the OMC transmitted
  signal is used. The quadrature sum is 16.1\,mrad rms.}
\begin{tabular}{l c}
\hline
source & rms phase [mrad]\\
\hline
in-loop: up to 2 kHz & 4 \\
audio: 2 kHz -- 45 kHz & 13 \\
RF: 14.9 MHz MI sidebands & 6.7 \\
RF: 9 MHz SRC sidebands & 5.5 \\
\hline
\end{tabular}
\label{tab:phasenoises}
\end{table}

Finally, Table~\ref{tab:phasenoises} presents a break down of the
known phase noise contributions from different frequency bands. The
audio band phase noise comes from computing the square difference
between the signal and the sensor noise in the OMC transmitted signal
from a measurement out to 45\,kHz. Above 45\,kHz, the signal is not
sensor-noise-limited. Phase noise from the 1\% of RF sidebands that
get transmitted through the OMC is computed \cite{Dwyer2013Squeezed}
based on a measurement of the power in the sidebands and the 5\% ratio
of contrast defect to dark fringe offset. Upon adding these
contributions incoherently, we find that we have a total known phase
noise of 16.1\,mrad rms. Phase noises which cannot be measured directly
such as phase noise in the frequency band above 45\,kHz and lock point
errors are not included. An out-of-loop measurement of phase noise
when using the OMC transmitted signal indicates the total phase noise
is $37\pm12$\,mrad rms.




\section{Discussion}
\label{sec:discussion}
Table~\ref{tab:properties} shows a summary of the various experimental
parameters that play a role in determining the quality of each of the
phase error signals. The most important new implications that this
work generates for interferometer design are related to the OMC and PD
readout electronics. The OMC finesse and the CCSB frequency can be
selected to allow a greater fraction of the CCSBs to be transmitted
through the OMC than the current 1\% transmission at GEO\,600. A
trade-off must be reached, however, between lowering the OMC finesse
to let more CCSBs through and preserving its function as a filter for
both HOMs and the interferometer RF sidebands. There are also
technical limitations to how low of a frequency the CCSBs can
be. Because the CCSBs have the same polarization as the squeezed
field, power noise on the CC field that extends into the GW frequency
band can reduce the squeezing generated at the OPA by seeding on top
of the vacuum seed. Some of this power noise could be alleviated by
stabilizing the CC laser.

\begin{table}
\centering
\caption{Important aspects to take into consideration when selecting
  which of the three squeezing angle error signals to use and when
  thinking about how to improve them.}
\begin{tabular}{l l l l}
\hline
 & Pick-off & OMC refl. & OMC trans. \\
\hline
HOMs (carrier) & $\times$ & &  \\
HOMs (MI SBs) &  & $\times$ & \\
HOMs (alignmnent) & $\times$ & $\times$ & \\
CCSB frequency & & & $\times$ \\
CCSB amplitude & $\times$ & $\times$ & $\times$ \\
MI SB amplitude & & $\times$ &  \\
Pick-off fraction & $\times$ & & \\
OMC finesse & & & $\times$ \\
\hline
\end{tabular}
\label{tab:properties}
\end{table}

Another aspect of using the OMC transmitted phase signal is that the
information is carried on the same light that carries the GW
signal. To maintain the highest possible detection efficiency, the two
signals must be detected using the exact same PD(s). While DC readout
of the GW signal requires only low-noise DC electronics, low-noise RF
electronics are needed in addition in order to recover the squeezer
phase error signal. This introduces the challenge of having to design
and build dual low noise DC and RF readout electronics that are also
not susceptible to RF pick-up. This represents current on-going work
at GEO\,600.

Although the OMC reflected phase signal is a good option for GEO\,600,
it is not necessarily the case for other detectors. The amount of MI
sideband HOMs will need to be evaluated for each individual
experimental setup. 
Also, the MI sidebands do intrinsically contribute
to phase noise at RF frequencies and a trade-off in the level of MI
sidebands is required. 
In addition, it should be noted that although the SNR argument based
on shot noise is irrelevant for GEO\,600 at the moment, it could be
meaningful in the future and for different detectors.


A nice side effect of eliminating the use of the pre-OMC pick-off for
a phase signal is that optical losses can be reduced. Although the
pick-off mirror is required for sensing some of the angular degrees of
freedom of the interferometer, a lower pick-off fraction can be
afforded when the light does not need to be shared with a PD for
squeezer phase sensing.

Finally, although we use a noise lock loop to counteract lock point
errors, it cannot fully compensate for all of the errors from the
pick-off signal due to its limited usable bandwidth of at most
100\,mHz. This limitation comes from the implementation of the noise
lock, which is to dither the squeezing phase at 11.6\,Hz. Higher
bandwidth could only come from increasing the dither amplitude, but
this itself would add to the rms phase noise. The noise lock loop is
thus limited to control unsensed drifts of the squeezing phase only on
slow time scales. Upon using the OMC transmitted signal, the noise
lock loop corrects for drifts of the squeezing angle on the order of
tens of mrad over hour time scales. One underlying cause of these
drifts is the fact that the CCSBs are imbalanced. Any change in the
relative amplitudes of the CCSBs results in an offset to the locking
point. This may arise from changes in non-linear gain which is itself
susceptible to influences such as changing laser power.

The phase signals in OMC reflection and OMC transmission have each
been used for standard squeezing operation at GEO\,600 at different
times since 2011. The greater part of the 11 month period reported in
Ref.~\cite{Grote2013First} used the OMC reflection signal, and since
the last couple of months of that run, the OMC transmitted signal has
been in permanent use. After a new signal recycling mirror was
installed at GEO\,600 which increased the amount of HOMs at the output
port and increased lock point errors, the use of these new signals was
a critical step for achieving stable squeezing.

\section{Conclusion}
\label{sec:conclusion}
We proposed two new methods of generating an error signal for matching
the longitudinal phase of squeezed states of vacuum to that of the
output field of a laser interferometer for gravitational-wave
detection. We experimentally compared both of the new methods to the
so-far standard method and the new methods are found to be superior.
As the main result of this work, we showed that squeezing phase
control using either of the new signals eliminates lock point errors
and greatly improves the squeezing level stationarity. We discussed
other features and advantages of the new methods which contribute to a
higher level of observed squeezing and considered some implications
for the design of future squeezed-vacuum applications. Having also
demonstrated the new methods in long-term application at GEO\,600, we
conclude that they are a pivotal development towards realizing stable
squeezing of 6\,dB or more in advanced detectors and beyond.

\section*{Acknowledgments}
The authors are grateful for support from the Science and Technology
Facilities Council (STFC), the University of Glasgow in the UK, the
Bundesministerium f{\"u}r Bildung und Forschung (BMBF), and the state
of Lower Saxony in Germany. This work was partly supported by the
Deutsche Forschungsgemeinschaft, DFG grant SFB/Transregio 7
Gravitational Wave Astronomy. Additionally, we thank Roman Schnabel
for leading the effort of providing the squeezer used at GEO\,600;
Alexander Khalaidovski, Christian Gr\"{a}f, and Nico Lastzka for their
work in building the squeezer and its automation system; Lisa Barsotti
and Sheila Dwyer for their respective visits and helpful input; Keita
Kawabe for his thoughtful comments on this manuscript; and Michael
Weinert, Volker Kringel, Marc Brinkmann and Walter Gra{\ss} for their
work in keeping the GEO\,600 interferometer in a good running state
for this experiment. This document has been assigned LIGO document
number LIGO-P1400150.


\end{document}